\documentclass[onecolumn]{IEEEtran}
\usepackage{graphics,color,subfigure,latexsym,amssymb,amsmath,url,amsthm}
\usepackage{bbm}
\usepackage[pdftex]{graphicx}

\newcommand{\eqdef} {\mathrel{\mathop:}=}
\newtheorem{definition}{Definition}
\newtheorem{remark}{Remark}
\newtheorem{theorem}{Theorem}

\newtheorem{lemma}{Lemma}
\newtheorem{proposition}{Proposition}

\newcommand{\be}{\begin{equation}}
\newcommand{\ee}{\end{equation}}
\newcommand{\ben}{\begin{enumerate}}
\newcommand{\een}{\end{enumerate}}
\newcommand{\bea}{\begin{eqnarray}}
\newcommand{\eea}{\end{eqnarray}}
\newcommand{\bean}{\begin{eqnarray*}}
\newcommand{\eean}{\end{eqnarray*}}

\newcommand{\cX}{{\cal X}}
\newcommand{\cY}{{\cal Y}}

\newcommand{\cA}{{\cal A}}

\newcommand{\cP}{{\cal P}}

\newcommand{\cS}{{\cal S}}

\newcommand{\cL}{{\cal L}}

\newcommand{\cM}{{\cal M}}

\def\eor{\hskip 3pt $\Diamond$}

\def\b0{\mathbf{0}}
\def\bX{\mathbf{X}}

\def\bY{\mathbf{Y}}

\def\bbR{\mathbb{R}}
\def\bbZ{\mathbb{Z}}
\def\bx{\mathbf{x}}
\def\by{\mathbf{y}}
\def\bz{\mathbf{z}}

\def\bb{\mathbf{b}}

\def\mE{\textnormal{E}}
\def\me{\textnormal{e}}

\def\mo{\textnormal{o}}
\def\mI{\textnormal{I}}

\def\mV{\textnormal{Var}}
\def\b1{\mathbbm{1}}
\def\W1{W_{1^{-},P,R}}

\def\mP{\textnormal{P}}

\def\mSP{\textnormal{SP}}

\def\eor{\hskip 3pt $\Diamond$}
\def\eot{\hskip 3pt $\blacklozenge$}
\def\eod{\hskip 3pt $\lozenge$}

\begin{document}

\title{The third-order term in the normal \\ approximation for singular channels}
\author{Y\"{u}cel Altu\u{g}~\IEEEmembership{Member }and Aaron B.~Wagner~\IEEEmembership{Senior Member}
\thanks{The authors are with the School of Electrical and Computer Engineering, Cornell University, 
Ithaca, NY 14853.
E-mail: {\em ya68@cornell.edu, wagner@ece.cornell.edu}.}}

\maketitle
\begin{abstract}
For a singular and symmetric discrete memoryless channel with positive dispersion, the third-order term in the normal approximation is shown to be upper bounded by a constant. This finding completes the characterization of the third-order term for symmetric discrete memoryless channels. The proof method is extended to asymmetric and singular channels with constant composition codes, and its connection to existing results, as well as its limitation in the error exponents regime, are discussed.    
\end{abstract}

\section{Introduction}
\label{sec:intro}
Decades after its introduction in information theory (e.g., \cite{juschkewtisch53}, \cite{strassen62}), the normal approximation has recently enjoyed a surge in interest. See, for instance, \cite{hayashi09}--\cite{tomamichel-tan13} for a partial list of recent work that is most closely related to the present paper. When particularized to coding over a discrete memoryless channel (DMC), say $W$, the normal approximation states that for any positive integer $n$ and $\epsilon \in (0,1)$, the logarithm of the maximum number of messages that can be communicated with an error probability not larger than $\epsilon$ behaves asymptotically as\footnote{Throughout the paper, we use nats as the unit of information.} 
\begin{equation}
n C(W) + \sqrt{n V_\epsilon(W)}\Phi^{-1}(\epsilon) + O(\ln n), 
\label{eq:intro-1}
\end{equation}
where $C(W)$ and $V_\epsilon(W)$ are the \emph{capacity} and the \emph{$\epsilon$-dispersion} of the channel, respectively, and $\Phi(\cdot)$ denotes the distribution of the standard normal random variable. 

Although the first two terms in \eqref{eq:intro-1} are well-understood, the third-order term has proven to be more elusive (e.g., \cite{hayashi09}, \cite[Sec.~3.4.5]{polyanskiy10-phd}, \cite{tomamichel-tan13}). Some bounds are available, however. The third-order term is known to be no greater than $\ln \sqrt{n}$~\cite{tomamichel-tan13} and no smaller than a constant (i.e., it cannot diverge to negative infinity)~\cite[Theorem~45]{polyanskiy10}. Each bound is tight for some channel.
The upper bound is tight for a class of channels that includes the binary symmetric channel (BSC)~\cite[Sec.~3.4.5]{polyanskiy10-phd}, \cite{tomamichel-tan13} while the lower bound is tight for the binary erasure channel (BEC)~\cite[Theorem~53]{polyanskiy10}. It is not known, however, whether these two extremes are the only possibilities.

In this paper, we prove that for symmetric channels\footnote{For a definition of symmetric channels, see Definition~\ref{def:symmetric}.}, these are indeed the only two possibilities. Specifically, we show that for a singular\footnote{For a definition of singular channels, see Definition~\ref{def:singularity}.} and symmetric DMC with positive dispersion, the third-order term is upper bounded by a constant (see Proposition~\ref{prop:sym}). By combining this finding with existing results in the literature, we can conclude that the third-order term for a symmetric DMC with positive dispersion is either $\ln \sqrt{n}$ or a constant depending whether the channel is nonsingular or singular (see Theorem~\ref{thrm:result} to follow).

It is worth noting that the analogous result for error exponents is already known~\cite{altug13}. For symmetric channels, the optimal order of the sub-exponential factor in the error exponents regime is $\Theta(n^{-0.5})$ in the singular case and $\Theta(n^{-0.5(1+|\mE^\prime(R)|)})$ in the nonsingular case, where $\mE^\prime(\cdot)$ is the derivative of the reliability function \cite{altug13}. In fact, the proof for the result presented here is based on the proof of this error exponent result. In Section~\ref{ssec:minmax}, we show how our main result can also be proven via the ``minimax converse'' (e.g., \cite[Theorem~1]{polyanskiy13}) in which a non-product output distribution is utilized. Polyanskiy~\cite[Section~VI.D]{polyanskiy13} had earlier showed how the constant upper bound on the third-order term in the normal approximation for the BEC could be obtained via the minimax converse with a particular non-product output distribution. Hence, an ancillary contribution of this paper is to show how the proof technique of Polyanskiy~\cite[Section~VI.D]{polyanskiy13} can be extended to all singular and symmetric channels.

Our proof technique can also be extended to asymmetric and singular channels, provided that attention is restricted to constant composition codes\footnote{A code is constant composition if all the codewords have the same empirical distribution.} (see Proposition~\ref{prop:asym}). In Section~\ref{ssec:CCC}, we discuss the difficulty in dropping this assumption for asymmetric and singular channels. 
Even with the constant composition assumption, our proof technique does not carry over easily to the error exponents regime if the channel is asymmetric, as we discuss in Section~\ref{ssec:EA}.


\section{Notation, definitions and statement of the results}
\label{sec:results}
\subsection{Notation}
\label{ssec:notation}
$\bbZ^+, \bbR, \bbR^{+}$ and $\bbR_{+}$ denote the set of positive integers, real, positive real and non-negative real numbers, respectively. Boldface letters denote vectors, regular letters with subscripts denote individual components of vectors. Furthermore, capital letters represent random variables and lowercase letters denote individual realizations of the corresponding random variable. For a finite set $\cX$, $\cP(\cX)$ (resp. $U_{\cX}$) denotes the set of all probability measures (resp. the uniform probability measure) on $\cX$. Similarly, for two finite sets $\cX$ and $\cY$, $\cP(\cY|\cX)$ denotes the set of all stochastic matrices from $\cX$ to $\cY$. Given any $P \in \cP(\cX)$, $\cS(P) \eqdef \{ x \in \cX \, : \, P(x) >0\}$. For any finite set $\cX$ and $n \in \bbZ^+$, $P_{\bx^n}$ denotes the type of the sequence $\bx^n$ and $\cP_n(\cX)$ denotes the set of all types on $\cX^n$. $\b1{\{ \cdot \} }$ denotes the standard indicator function. $\phi(\cdot)$ denotes the density of the standard Gaussian random variable. For a set $\cS$, $\textnormal{cl}(S)$ and $\cS^{c}$ denotes the closure of $\cS$ and the complementary set, respectively. We follow the notation of the book of Csisz\'ar-K\"orner \cite{csiszar-korner81} for standard information theoretic quantities. 

\subsection{Definitions}
\label{ssec:definitions}
\begin{definition}
For any $n \in \bbZ^+$ and $\epsilon \in (0,1)$,
\begin{equation}
M^\ast(n, \epsilon) \eqdef \max\{\lceil e^{nR} \rceil \in \bbR_+ : \bar{\mP}_{\me}(n,R) \leq \epsilon \}, \label{eq:Mstar}
\end{equation}
where $\bar{\mP}_{\me}(n,R)$ denotes the minimum average error probability attainable by any $(n,R)$ code. Further, for any $n \in \bbZ^+$ and $\epsilon \in (0,1)$,
\begin{equation}
M^\ast_{\textnormal{c}}(n, \epsilon) \eqdef \max\{\lceil e^{nR} \rceil \in \bbR_+ : \bar{\mP}_{\me, \textnormal{c}}(n,R) \leq \epsilon \}, \label{eq:Mstar-c}
\end{equation}
where $\bar{\mP}_{\me, \textnormal{c}}(n,R)$ denotes the minimum average error probability attainable by any $(n,R)$ constant composition code.\eod
\label{def:Mstar}
\end{definition}

\begin{definition}(Gallager~\cite[pg.~94]{gallager68})
A discrete memoryless channel is \emph{symmetric} if the channel outputs can be partitioned into subsets such that within each subset, the matrix of transition probabilities satisfies the following: every row (resp. column) is a permutation of every other row (resp. column). A channel that is not symmetric is called \emph{asymmetric}. \eod
\label{def:symmetric}
\end{definition}

\begin{definition}(\cite[Definition~2]{altug13})
A channel $W \in \cP(\cY|\cX)$ is \emph{singular} if 
\begin{equation}
\forall \, (x,y,z) \in \cX \times \cY \times \cX \textnormal{ with } W(y|x)W(y|z)>0, \, W(y|x)=W(y|z).
\label{eq:singularity}
\end{equation}
A channel that is not singular is called \emph{nonsingular}. \eod
\label{def:singularity}
\end{definition}

Given $W \in \cP(\cY|\cX)$, $C(W)$ denotes the capacity of the channel. For any $P \in \cP(\cX)$, define $q_P(y) \eqdef \sum_{x \in \cX}P(x)W(y|x)$. For convenience, let $q$ denote $q_{U_{\cX}}$. Given any $W \in \cP(\cY|\cX)$, $P \in \cP(\cX)$ and $\epsilon \in (0,1)$, define (e.g., \cite[Sec.~3.4]{polyanskiy10-phd})
\begin{align}
V(P,W) & \eqdef \sum_{x,y}P(x)W(y|x) \left[  \ln \frac{W(y|x)}{q_P(y)} -  \sum_{b}W(b|x)\ln\frac{W(b|x)}{q_P(b)}\right]^2. \label{eq:dispersion}\\
V_\epsilon(W) & \eqdef 
\begin{cases}
\min_{Q : \mI(Q;W) = C(W)}V(Q,W), & \textnormal{ if } \epsilon \in (0,1/2), \\
\max_{Q : \mI(Q;W) = C(W)}V(Q,W), & \textnormal{ if } \epsilon \in [1/2,1).
\end{cases} \label{eq:eps-dispersioin}\\
V^r(P,W) & \eqdef \sum_{x,y}P(x)W(y|x) \left[  \ln \frac{W(y|x)}{q_P(y)} - \sum_{z}\frac{P(z)W(y|z)}{q_P(y)}\ln\frac{W(y|z)}{q_P(y)}\right]^2. \label{eq:reverse-dispersion}
\end{align}

The following result gives an equivalent definition of singularity in terms of the quantity defined in \eqref{eq:reverse-dispersion}.

\begin{lemma}
Consider a channel $W$ and $P \in \cP(\cX)$ with $P(x) >0$ for all $x \in \cX$. $V^r(P, W)=0$ if and only if $W$ is singular. \eot
\label{lem:singular}
\end{lemma}

\begin{IEEEproof}
We note that 
\begin{equation}
\left[ V^r(P,W) = 0 \right] \Longleftrightarrow \left[ \forall \, y \in \cY, \, \ln W(y|x) = \sum_{z}\frac{P(z)W(y|z)}{q_P(y)}\ln W(y|z), \, \forall x \in \cX \textnormal{ with } W(y|x)>0 \right], \label{eq:prelim-pf1}
\end{equation}
which is a direct consequence of the definition of $V^r(U_\cX,W)$, i.e., \eqref{eq:reverse-dispersion}. In light of Definition~\ref{def:singularity}, the right side of \eqref{eq:prelim-pf1} is equivalent to saying $W$ is singular. 
\end{IEEEproof}

\begin{remark}
In \cite[Lemma~52]{polyanskiy10-phd}, it is claimed that
\begin{equation}
\left[ V^r(P,W) = 0 \right] \Longleftrightarrow \left[ \forall \, (x,y,y^\prime) \, : \, W(y|x)=W(y^\prime|x) \textnormal{ or } P(x)W(y|x)=0 \right].
\label{eq:rem-yury}
\end{equation}
By choosing $P = U_\cX$ and $W$ as BEC with parameter $\delta \in (0,1)$, one can verify that $V^r(P,W) = 0$ by elementary calculation. Evidently, this $(P,W)$ pair does not satisfy the right side of \eqref{eq:rem-yury} and hence \eqref{eq:rem-yury} is incorrect. 

If one replaces the right side of \eqref{eq:rem-yury} with the following modified definition of singularity\footnote{Note that \eqref{eq:rem-yury-1} is the definition of singularity given in \cite[Definition~1]{altug13-1}.}
\begin{equation}
\forall \, (x,y,z) \in \cX \times \cY \times \cX \textnormal{ with } P(x)W(y|x)P(z)W(y|z) >0, \, W(y|x) = W(y|z), \label{eq:rem-yury-1}
\end{equation}
then, by noticing 
\begin{equation}
\left[ V^r(P,W) =0 \right] \Longleftrightarrow \left[ \forall \, y \in \cY, \, \ln W(y|x) = \sum_{z}\frac{P(z)W(y|z)}{q_P(y)}\ln W(y|z), \, \forall \, x \in \cX  \textnormal{ with } P(x)W(y|x) >0 \right], 
\end{equation}
it is easy to see that \eqref{eq:rem-yury} holds. \eor 
\label{rem:rem-yury}
\end{remark}

\subsection{Results}
\label{ssec:result}
\begin{proposition}
Given $\epsilon \in (0,1)$ and a singular, symmetric $W$ with $V_\epsilon(W)>0$, we have 
\begin{equation}
\ln M^\ast(n, \epsilon) \leq n C(W) + \sqrt{n V_\epsilon(W)}\Phi^{-1}(\epsilon) + K(\epsilon,W), 
\label{eq:prop-sym}
\end{equation}
where $K(\epsilon, W) \in \bbR^+$ is a constant that depends on $\epsilon$ and $W$. \eot
\label{prop:sym}
\end{proposition}
\begin{IEEEproof}
Given in Section~\ref{ssec:sym}.
\end{IEEEproof}

Proposition~\ref{prop:sym} completes the proof of the following theorem. 
\begin{theorem}
Given a symmetric $W$ and $\epsilon \in (0,1)$, we have the following:
\begin{itemize}
\item[(i)] If $W$ is nonsingular and $V_\epsilon(W)>0$, then 
\begin{equation}
\ln M^\ast(n,\epsilon) = nC(W) + \sqrt{n V_\epsilon(W)}\Phi^{-1}(\epsilon) + \ln\sqrt{n} + \Theta(1). 
\label{eq:nonsingular}
\end{equation}
\item[(ii)] If $W$ is singular and $V_\epsilon(W) >0$, then 
\begin{equation}
\ln M^\ast(n,\epsilon) = nC(W) + \sqrt{n V_\epsilon(W)}\Phi^{-1}(\epsilon) + \Theta(1). 
\label{eq:singular}
\end{equation}
\item[(iii)] If $V_\epsilon(W)=0$, then 
\begin{equation}
\ln M^\ast(n,\epsilon) = nC(W) + \Theta(1). 
\label{eq:peculiar}
\end{equation}
\end{itemize}
\label{thrm:result}
\eot
\end{theorem}
\begin{IEEEproof}
We point out the existing results that justify the cases except the converse statement of item (ii), which follows from Proposition~\ref{prop:sym}. Achievability of item (i) follows from \cite[Corollary~54]{polyanskiy10-phd} that is applicable due to Lemma~\ref{lem:singular}. Converse of item (i) follows from \cite[Theorem~55]{polyanskiy10-phd}. Achievability of item (ii) follows from \cite[Theorem~47]{polyanskiy10-phd}, coupled with item (ii) of Lemma~\ref{lem:sym-prelim}. Item (iii) is proved in \cite[Corollary~57]{polyanskiy10-phd}. 
\end{IEEEproof}

\begin{proposition}
Given a singular and asymmetric $W$, we have  
\begin{itemize}
\item[(i)] If $\epsilon \in (0,1/2)$, then 
\begin{equation}
\ln M^\ast_c(n, \epsilon) \leq n C(W) + \sqrt{n V_\epsilon(W)}\Phi^{-1}(\epsilon) + \tilde{K}(\epsilon,W), 
\label{eq:prop-asym-1}
\end{equation}
where $\tilde{K}(\epsilon, W) \in \bbR^+$ is a constant that depends on $\epsilon$ and $W$.
\item[(ii)] If $\epsilon \in (1/2,1)$ and $V_\epsilon(W) >0$, then 
\begin{equation}
\ln M^\ast_c(n, \epsilon) \leq n C(W) + \sqrt{n V_\epsilon(W)}\Phi^{-1}(\epsilon) + \tilde{K}^\prime(\epsilon,W), \label{eq:prop-asym-2}
\end{equation}
where $\tilde{K}^\prime(\epsilon, W) \in \bbR^+$ is a constant that depends on $\epsilon$ and $W$. \eot
\end{itemize}
\label{prop:asym}
\end{proposition}
\begin{IEEEproof}
Given in Section~\ref{ssec:asym}.
\end{IEEEproof}

\begin{remark}
\begin{itemize}
\item[(i)] The set of asymmetric and singular channels is not empty. For an example, let $\cX  \eqdef \{ 0,1,2\}$, $\cY \eqdef \{ 0, 1, 2, 3\}$ and consider 
\begin{equation}
W(y|x) \eqdef  
\begin{cases}
2/3, & \textnormal{ if } (x,y) = (0,0) \\
1/6, & \textnormal{ if } (x,y) \in \{ (0,1), (0, 3), (1,3), (2,1) \} \\
5/6, & \textnormal{ if } (x,y) \in \{ (1, 2), (2, 2)\} \\
0, & \textnormal{ else}. 
\end{cases}
\end{equation}
\item[(ii)] We do not analyze the zero-dispersion case for $\epsilon \in (1/2,1)$, because the third-order term also depends on whether the channel is \emph{exotic} (e.g., \cite[pg.~2331]{polyanskiy10}) and the main purpose of this paper is to investigate the effect of singularity on the third-order term. Similarly, we do not consider $\epsilon = 1/2$ case, since the third-order term also depends on whether the channel is exotic. See \cite[Section III]{tomamichel-tan13} for a detailed discussion on the effect of the exotic property of the channel on the third-order term when $\epsilon \in [1/2,1)$. \eor
\end{itemize}
\end{remark}

\section{Proofs}
\label{sec:proof}
First, we prove two lemmas that will be used in the proofs of both Proposition~\ref{prop:sym} and \ref{prop:asym}. To this end, consider any $Q \in \cP(\cX)$ and define 
\begin{equation}
\alpha_y(Q) \eqdef \sum_{x:W(y|x)>0}Q(x).
\label{eq:alpha}
\end{equation}
Consider any singular $W \in \cP(\cY|\cX)$. As a direct consequence of the singularity of the channel, for any $y \in \cY$, $W(y|x)$ is either $0$ or a column specific positive constant, say $\delta_y$. For any $y \in \cY$, $q_Q(y) = \delta_y \alpha_y(Q)$. The following set will be instrumental in our analysis:
\begin{equation}
\cS_R(Q) \eqdef \left\{ \by^n : \frac{1}{n}\sum_{i=1}^n \ln \frac{1}{\alpha_{y_i}(Q)} \leq R \right\}, 
\label{eq:SR}
\end{equation}
for any $R \in \bbR_+$. 

\begin{lemma}
Consider a singular $W \in \cP(\cY|\cX)$. Consider any $(n,R)$ code, say $(f, \varphi)$, with codewords $\{ \bx^n(m)\}_{m=1}^{|\cM|}$, where $\cM \eqdef \{ 1, \ldots, \lceil e^{nR}\rceil\}$ denotes the set of messages. Let $\bar{\mP}_{\me}(f, \varphi)$ denote the average error probability of this code. Fix some $Q \in \cP(\cX)$ and $\bz^n \in \cX^n$ and assume that for all $m \in \cM$, $W(\cS_R(Q)|\bx^n(m)) = W(\cS_R(Q) | \bz^n)$ and $q_Q$ dominates $W(\cdot|x)$ for all $x \in \cS(P_{\bx^n(m)})$. Then we have 
\begin{equation}
\bar{\mP}_{\me}(f, \varphi) \geq W(\cS_R(Q) | \bz^n) - \sum_{\by^n \in \cS_R(Q)}q_Q(\by^n) e^{-n\left[ R - \frac{1}{n}\sum_{i=1}^n \ln \frac{1}{\alpha_{y_i}(Q)}\right]}. 
\label{eq:lem-multipurp}
\end{equation}
\eot
\label{lem:multipurp}
\end{lemma}
\begin{IEEEproof}
Assume $W(\cS_R(Q)|\bz^n) >0$, otherwise \eqref{eq:lem-multipurp} is trivial. For any $\bx^n \in \cX^n$ with $W(\cS_R(Q) | \bx^n) >0$, define 
\begin{equation}
P_{Y|X, \cS_R(Q)}(\by^n | \bx^n , \cS_R(Q)) \eqdef 
\begin{cases}
\frac{W(\by^n | \bx^n )}{W( \cS_R(Q) | \bx^n)}, & \textnormal{ if } \by^n \in \cS_R(Q), \\
0, & \textnormal{ else}. 
\end{cases}
\label{eq:lem-multipurp-cond-dist}
\end{equation}
Evidently, $P_{Y|X, \cS_R(Q)}(\cdot | \bx^n , \cS_R(Q)) $ is a well-defined probability measure. Let $\{ \mathcal{A}_m\}_{m=1}^{|\cM|}$ denote the decoding regions of the code. We have 
\begin{align}
\bar{\mP}_{\me}(f, \varphi) & = \frac{1}{|\cM|}\sum_{m \in \cM} \sum_{\by^n \in \cA_m^c}W(\by^n | \bx^n(m))  \\
& \geq \frac{1}{|\cM|}\sum_{m \in \cM} \sum_{\by^n \in \cA_m^c} W(\cS_R(Q) | \bx^n(m)) P_{Y|X, \cS_R(Q)}(\by^n | \bx^n(m), \cS_R(Q)) \label{eq:lem-multipurp-pf1} \\
& \geq W(\cS_R(Q) | \bz^n) \left[1 - \frac{1}{|\cM|}\sum_{m \in \cM} \sum_{\by^n \in \cA_m}  P_{Y|X, \cS_R(Q)}(\by^n | \bx^n(m), \cS_R(Q)) \right]\label{eq:lem-multipurp-pf2}, 
\end{align}
where \eqref{eq:lem-multipurp-pf1} follows from \eqref{eq:lem-multipurp-cond-dist} and \eqref{eq:lem-multipurp-pf2} follows from the assumption that $W(\cS_R(Q)|\bx^n(m)) = W(\cS_R(Q) | \bz^n)$, for all $m \in \cM$. Define $P_{D|Y}(m| \by^n) \eqdef \b1\{ \by^n \in \cA_m\}$, for all $m \in \cM$. Since the decoding regions are mutually exclusive and collectively exhaustive on $\cM$, $P_{D|Y}(\cdot| \by^n) $ is a well-defined probability measure. Hence, \eqref{eq:lem-multipurp-pf2} implies that 
\begin{align}
\bar{\mP}_{\me}(f, \varphi) & \geq W(\cS_R(Q) | \bz^n) \left[ 1 - \frac{e^{-nR}}{W(\cS_R(Q)| \bz^n)} \sum_{m \in \cM} \sum_{\by^n} P_{D|Y}(m|\by^n) W(\by^n | \bx^n(m)) \b1\{ \by^n \in \cS_R(Q)\}\right]  \\
& \geq  W(\cS_R(Q) | \bz^n) \left[ 1 - \frac{e^{-nR}}{W(\cS_R(Q)| \bz^n)} \sum_{m \in \cM} \sum_{\by^n} P_{D|Y}(m|\by^n)  \b1\{ \by^n \in \cS_R(Q)\} q_Q(\by^n)e^{\sum_{i=1}^n \ln \frac{1}{\alpha_{y_i}(Q)}}  \right] \label{eq:lem-multipurp-pf3}\\
& \geq W(\cS_R(Q) | \bz^n) \left[ 1 - \frac{e^{-nR}}{W(\cS_R(Q)| \bz^n)}  \sum_{\by^n} \b1\{ \by^n \in \cS_R(Q)\} q_Q(\by^n)e^{\sum_{i=1}^n \ln \frac{1}{\alpha_{y_i}(Q)}}  \right],  
\end{align}
where \eqref{eq:lem-multipurp-pf3} follows from the fact that $q_Q(y) = \delta_y \alpha_y(Q)$ and the assumption that for all $m \in \cM$, $q_Q$ dominates $W(\cdot|x)$ for all $x \in \cS(P_{\bx^n(m)})$.
\end{IEEEproof}

\begin{remark}
Lemma~\ref{lem:multipurp} has the following intuitive interpretation. For simplicity, consider an $(n,R)$ constant composition code $(f, \varphi)$ with the common composition $Q$. We write  
\begin{equation}
\bar{\mP}_{\me}(f, \varphi) = \Pr(\cS_R(Q)) \bar{\mP}_{\me}(f, \varphi | \cS_R(Q)),
\label{eq:rem-lemma1}
\end{equation}
where $\bar{\mP}_{\me}(f, \varphi | \cS_R(Q))$ denotes the average error probability of $(f,\varphi)$ conditioned on $\cS_R(Q)$. 

Given any $\bx^n \in \cX^n$ with $P_{\bx^n} = Q$, $\cS_R(Q)$ captures the event that the empirical mutual information, i.e., $\frac{1}{n}\sum_{i=1}^n \ln \frac{W(Y_i|x_i)}{q_Q(Y_i)}$, is smaller than $R$ as a direct consequence of the singularity of $W$. Intuitively, the code will make an error if the channel realization is such that the resulting empirical mutual information is not large enough to support the coding rate, and this is our rationale in writing \eqref{eq:rem-lemma1}. Since $(f, \varphi)$ is a constant composition code, one can write $W(\cS_R(Q) | \bx^n)$ in place of $\Pr(\cS_R(Q))$ and 
\begin{equation}
1 - \frac{1}{W(\cS_R(Q) | \bx^n)}\sum_{\by^n \in \cS_R(Q)}q_Q(\by^n) e^{-n\left[ R - \frac{1}{n}\sum_{i=1}^n \ln \frac{1}{\alpha_{y_i}(Q)}\right]}
\end{equation}
can be viewed as a lower bound on $\bar{\mP}_{\me}(f, \varphi | \cS_R(Q))$. Therefore, \eqref{eq:lem-multipurp} can be considered as a lower bound on the right side of \eqref{eq:rem-lemma1}. 
\eor
\label{rem:lemma1}
\end{remark}

We continue with a simple result for sums of independent random variables whose proof is inspired by the proof of \cite[Lemma~47]{polyanskiy10}. The reason of its inclusion is the fact that the bound in \eqref{eq:lem2-1} is tighter than the one that follows by a direct application of \cite[Lemma~47]{polyanskiy10}, at least by a factor of $2$.
\begin{lemma}
Let $\{ Z_i\}_{i=1}^n$ be independent with $m_{2,n} \eqdef \sum_{i=1}^n \mV[Z_i] >0$ and $m_{3,n} \eqdef \sum_{i=1}^n \mE\left[ | Z_i - \mE\left[ Z_i\right]|^3 \right] < \infty$. Then for any $r \in \bbR$ and $n \in \bbZ^+$ 
\begin{equation}
\mE\left[   \b1\left\{ \sum_{i=1}^n Z_i  \leq r \right\} e^{-\left[ r - \sum_{i=1}^n Z_i\right] }\right] \leq \frac{1}{\sqrt{2 \pi m_{2,n}}} + \frac{2 m_{3,n}}{m_{2,n}^{3/2}}. \label{eq:lem2-1}
\end{equation}
Further, if the random variables are also identically distributed, then 
\begin{equation}
\mE\left[   \b1\left\{ \sum_{i=1}^n Z_i  \leq r \right\} e^{-\left[ r - \sum_{i=1}^n Z_i\right] }\right] \leq \frac{1}{\sqrt{2 \pi m_{2,n}}} + \frac{m_{3,n}}{m_{2,n}^{3/2}}. \label{eq:lem2-2}
\end{equation}
\label{lem:lem2}
\eot
\end{lemma}
\begin{IEEEproof}
Define $ S_n \eqdef \sum_{i=1}^n Z_i$ and let $F_n$ denote the distribution function of $S_n$. For convenience, let $B_n(r)$ denote the left side of \eqref{eq:lem2-1} and $m_{1,n} \eqdef \sum_{i=1}^n \mE[Z_i]$. We have 
\begin{align}
B_n(r) & = e^{-r} \int_{-\infty}^{r} e^{z} dF_n(z)  \\
& = F_n(r) - \int_{-\infty}^{r}e^{(z - r)}F_n(z)dz \label{eq:lem2-pf1}\\
& = \int_{0}^{\infty} e^{-x}\left[ F_n(r) - F_n\left(r-x\right) \right] dx  \\
& \leq \int_{0}^{\infty}e^{-x}\left\{ \int_{\frac{r - m_{1,n}}{\sqrt{m_{2,n}}} - \frac{x}{\sqrt{m_{2,n}}}}^{\frac{r - m_{1,n}}{\sqrt{m_{2,n}}}} \frac{e^{-\frac{a^2}{2}}}{\sqrt{2 \pi}} da + c\frac{ m_{3,n}}{m_{2,n}^{3/2}}\right\}dx  \label{eq:lem2-pf2}\\
& \leq \frac{1}{\sqrt{2 \pi m_{2,n} }} + c\frac{ m_{3,n}}{m_{2,n}^{3/2}} ,  
\end{align}
where \eqref{eq:lem2-pf1} follows from integration by parts, \eqref{eq:lem2-pf2} follows from the Berry-Esseen theorem\footnote{We take the constant in the Berry-Esseen theorem as $1$ (resp. $1/2$) if the random variables are independent (resp. i.i.d.), although neither choice is the best possible (e.g., \cite{korolev2010}).} and $c=2$ (resp. $c=1$) if the random variables are independent (resp. i.i.d.). 
\end{IEEEproof}

\subsection{Proof of Proposition~\ref{prop:sym}}
\label{ssec:sym}
Let $W \in \cP(\cY | \cX)$ be a symmetric and singular channel. Without loss of generality, assume $W$ has no all-zero column. Consider any $\epsilon \in (0,1)$. Define 
\begin{equation} 
\forall \, x \in \cX, \, M_x(\lambda) \eqdef \mE_{W(\cdot|x)}\left[ e^{\lambda \ln \frac{W(Y|x)}{q(Y)}}\right], \, m_3(x) \eqdef \mE_{W(\cdot|x)}\left[ \left| \ln \frac{W(Y|x)}{q(Y)} - C(W) \right|^3\right], 
\end{equation} 
for any $\lambda \in \bbR$. For convenience, let $\cS_R$ denotes $\cS_R(U_\cX)$, which is defined in \eqref{eq:SR}. 

\begin{lemma} Let $W \in \cP(\cY|\cX)$ be a symmetric and singular channel. Fix an arbitrary $x_\mo \in \cX$. 
\begin{itemize}
\item[(i)] For any $x \in \cX$, $M_x(\lambda) = M_{x_{\mo}}(\lambda)$ for all $\lambda \in \bbR$. 

\item[(ii)] For all $x \in \cX$, 
\begin{align}
& \mE_{W(\cdot|x)}\left[ \ln \frac{W(Y|x)}{q(Y)}\right] = \mE_{W(\cdot|x_\mo)}\left[ \ln \frac{W(Y|x_\mo)}{q(Y)}\right] = C(W), \label{eq:lem-sym-prelim-1} \\
& \mV_{W(\cdot|x)}\left[ \ln \frac{W(Y|x)}{q(Y)}\right] = \mV_{W(\cdot|x_\mo)}\left[ \ln \frac{W(Y|x_\mo)}{q(Y)}\right] =: V(W) = V_\epsilon(W), \label{eq:lem-sym-prelim-2}\\
& m_3(x) = m_3(x_\mo). \label{eq:lem-sym-prelim-3}
\end{align}
 
\item[(iii)] For any $\bx^n \in \cX^n$, $W(\cS_R | \bx^n) = W(\cS_R | \bx_\mo^n)$, where $\bx^n_\mo$ denotes the element of $\cX^n$ consisting of all $x_\mo$.   

\item[(iv)] $\mE_{q}[-\ln \alpha_Y] = C(W), \mV[-\ln \alpha_Y] = V(W)$ and $\mE_q[|-\ln \alpha_Y - C(W)|^3] = m_3(x_\mo)$. \eot
\end{itemize}
\label{lem:sym-prelim}
\end{lemma}
\begin{IEEEproof}
Since $U_\cX$ is a capacity achieving input distribution of $W$ (e.g., \cite[Theorem~4.5.2]{gallager68}) and the unique capacity achieving output distribution has full support (e.g., \cite[Corollary 1 and 2 to Theorem~4.5.1]{gallager68}), we conclude that $\alpha_y >0$, for all $y \in \cY$.  

\begin{itemize}
\item[(i)] Let $\{ \cY_l \}_{l=1}^L$ be a partition\footnote{The choice of the partition will be immaterial in the sequel.} mentioned in Definition~\ref{def:symmetric}. Let $\{ W_{l}\}_{l=1}^{L}$ denote the sub-channel associated with each $\cY_l$, which is simply the matrix formed by using the columns of $\cY_l$. Evidently, for any $l \in \{ 1, \ldots, L \} =: \mathcal{L}$ and $y_1, y_2 \in \cY_l$, $\delta_{y_1} = \delta_{y_2}$ and hence, for any $l \in \mathcal{L}$, any entry of $W_l$ can take only two values, either $0$ or $\delta_l$ with $\delta_l \eqdef \delta_y$ for some $y \in \cY_l$. Following similar reasoning, $\alpha_{(\cdot)}$ is also constant along $\cY_l$ and we define $\alpha_l \eqdef \alpha_y$ for all $y \in \cY_l$. For any $l \in \mathcal{L}$, let $\nu_{l}$ denote the number of positive elements in a row of $W_{l}$. 

For any $x \in \cX$ and $\lambda \in \bbR$, we have 
\begin{equation}
M_x(\lambda) = \sum_{y: W(y|x) >0}\delta_y \alpha_y^{-\lambda} = \sum_{l \in \cL}\nu_l \delta_l \alpha_l^{-\lambda},
\label{eq:lem-sym-prelim-pf1}
\end{equation}
where the second equality follows due to the symmetry of the channel. Evidently, \eqref{eq:lem-sym-prelim-pf1} ensures that $M_x(\cdot)$ is finite on $\bbR$ and also implies item (i). 

\item[(ii)] Item (i), along with the uniqueness theorem for moment generating functions (e.g., \cite[Ex.~26.7]{billingsley95}), directly implies \eqref{eq:lem-sym-prelim-1}, \eqref{eq:lem-sym-prelim-3} and 
\begin{equation}
\mV_{W(\cdot|x)}\left[ \ln \frac{W(Y|x)}{q(Y)}\right] = V(W), \forall \, x \in \cX. 
\label{eq:lem-sym-prelim-pf2}
\end{equation}
The last equality of \eqref{eq:lem-sym-prelim-2} is evident in light of \eqref{eq:lem-sym-prelim-pf2} and the fact that $q$ is the unique capacity achieving output distribution of $W$. 

\item[(iii)] The claim is a direct consequence of item (i) of this lemma. 

\item[(iv)] The claim directly follows from item (ii) of this lemma on account of the definition of $q$ and the fact that $q(y) = \delta_y \alpha_y$. 

\end{itemize}
\end{IEEEproof}

\begin{remark}  
Equations \eqref{eq:lem-sym-prelim-1} and \eqref{eq:lem-sym-prelim-2} are proved in \cite[Theorem~55]{polyanskiy10-phd} for the set of weakly input symmetric channels that subsumes symmetric channels. \eor
\end{remark}

We conclude the proof as follows. First, we define 
\begin{align}
k(W) & \eqdef \frac{m_3(x_\mo)}{V(W)^{3/2}}, \\ 
K(\epsilon, W) & \eqdef \frac{k(W)\sqrt{V(W)}}{\phi(\Phi^{-1}(\epsilon))} + \frac{2}{\phi(\Phi^{-1}(\epsilon))}\left( \frac{1}{\sqrt{2\pi}} + \frac{m_3(x_\mo)}{V(W)}\right). \label{eq:sym-pf-final0}
\end{align}
Evidently, $K(\epsilon, W) \in \bbR^+$. Choose some $n_\mo(\epsilon, W) \in \bbZ^+$ such that for all $n \geq n_\mo(\epsilon,W)$, 
\begin{equation}
\left\{ 1 - \frac{K(\epsilon,W)}{2 \phi(\Phi^{-1}(\epsilon))\sqrt{n V(W)}}\right\} > 1/2.
\label{eq:sym-pf-final1}
\end{equation}
Consider any $n \geq n_\mo(\epsilon,W)$ and define 
\begin{equation}
R \eqdef C(W) + \sqrt{\frac{V(W)}{n}}\Phi^{-1}(\epsilon) + \frac{K(\epsilon,W)}{n}. 
\label{eq:sym-pf-final2} 
\end{equation}
Consider any $(n,R)$ code, say $(f, \varphi)$. Due to the fact that $q(y)>0$ for all $y \in \cY$ and item (iii) of Lemma~\ref{lem:sym-prelim}, we can apply Lemma~\ref{lem:multipurp} to deduce 
\begin{equation}
\bar{\mP}_{\me}(f, \varphi) \geq W(\cS_R | \bx^n_{\mo}) - \sum_{\by^n \in \cS_R}q(\by^n)e^{-n\left[R - \frac{1}{n}\sum_{i=1}^n \ln \frac{1}{\alpha_{y_i}}\right]}. \label{eq:sym-pf-final3}
\end{equation}
Since $V(W)>0$, item (iv) of Lemma~\ref{lem:sym-prelim} ensures that we can apply Lemma~\ref{lem:lem2} to have 
\begin{equation}
\sum_{\by^n \in \cS_R}q(\by^n)e^{-n\left[R - \frac{1}{n}\sum_{i=1}^n \ln \frac{1}{\alpha_{y_i}}\right]} \leq \frac{1}{\sqrt{2 \pi n V(W)}} + \frac{k(W)}{\sqrt{n}}. \label{eq:sym-pf-final4}
\end{equation}
Next, we claim that 
\begin{equation}
W(\cS_R | \bx^n_\mo) \geq \epsilon + \frac{K(\epsilon,W)\phi(\Phi^{-1}(\epsilon))}{\sqrt{nV(W)}}\left\{ 1 - \frac{K(\epsilon,W)}{\phi(\Phi^{-1}(\epsilon))2 \sqrt{n V(W)}}\right\} - \frac{k(W)}{2 \sqrt{n}}. 
\label{eq:sym-pf-final5}
\end{equation}
To see \eqref{eq:sym-pf-final5}, we note that 
\begin{align}
W(\cS_R | \bx^n_\mo) & = W\left\{  \frac{1}{n} \sum_{i=1}^n \ln \frac{W(Y_i|x_\mo)}{q(Y_i)} \leq R \, \bigg| \, \bx^n_\mo \right\} \label{eq:sym-pf-final5.1} \\
& = W\left\{\frac{1}{\sqrt{n V(W)}}\sum_{i=1}^n \left[ \ln \frac{W(Y_i|x_\mo)}{q(Y_i)} - C(W) \right] \leq \Phi^{-1}(\epsilon) + \frac{K(\epsilon,W)}{\sqrt{n V(W)}} \, \bigg| \, \bx^n_\mo \right\} \label{eq:sym-pf-final5.2} \\
& \geq \Phi\left( \Phi^{-1}(\epsilon) + \frac{K(\epsilon, W)}{\sqrt{n V(W)}}\right) - \frac{k(W)}{2\sqrt{n}}, \label{eq:sym-pf-final5.3}
\end{align}
where \eqref{eq:sym-pf-final5.1} follows since $q(y) = \delta_y \alpha_y$, along with the singularity of the channel, \eqref{eq:sym-pf-final5.2} follows from the definition of $R$, i.e., \eqref{eq:sym-pf-final2}, and \eqref{eq:sym-pf-final5.3} follows from the Berry-Esseen theorem, whose applicability is ensured by item (ii) of Lemma~\ref{lem:sym-prelim} and the fact that $V(W)>0$. Via a straightforward power series approximation, one can check that \eqref{eq:sym-pf-final5.3} implies \eqref{eq:sym-pf-final5}.

By plugging \eqref{eq:sym-pf-final4} and \eqref{eq:sym-pf-final5} into \eqref{eq:sym-pf-final3}, along with \eqref{eq:sym-pf-final1} and noticing the fact that the code is arbitrary, we deduce that 
\begin{equation}
\forall \, n \geq n_\mo(\epsilon, W), \, \ln M^\ast(n, \epsilon) \leq nC(W) + \sqrt{n V(W)}\Phi^{-1}(\epsilon) + K(\epsilon, W),
\end{equation}
which, in turn, implies the desired result. \qed

\subsection{Proof of Proposition~\ref{prop:asym}}
\label{ssec:asym}
We separately analyze three different possibilities for the composition of the code $P$: large $\mI(P;W)$ with large $V(P,W)$, large $\mI(P;W)$ with small $V(P,W)$, and small $\mI(P;W)$. This idea originated in Strassen's classical paper~\cite{strassen62} and is frequently used in the normal approximation regime. 

Specifically, given any $\delta, \nu \in \bbR^+$, we define 
\begin{align}
\cS_1(\delta, \nu) & \eqdef  \left\{P \in \cP(\cX) : \min_{P^\ast \in \cP^\ast_W} || P - P^\ast||_2 \leq \delta \textnormal{ and } V(P,W) \geq \nu \right\}, \label{eq:S1}\\
\cS_2(\delta, \nu) & \eqdef  \left\{P \in \cP(\cX) : \min_{P^\ast \in \cP^\ast_W} || P - P^\ast||_2 \leq \delta \textnormal{ and } V(P,W) < \nu \right\}, \label{eq:S2}\\
\cS_3(\delta) & \eqdef \left\{P \in \cP(\cX) : \min_{P^\ast \in \cP^\ast_W} || P - P^\ast||_2 > \delta  \right\}, \label{eq:S3}
\end{align}
where $\cP^\ast_W \eqdef \{ P \in \cP(\cX) : \mI(P;W) = C(W) \}$. Throughout the section, $\bar{\mP}_{\me}(f, \varphi)$ denotes the average error probability of the code $(f,\varphi)$.

\begin{lemma}
Fix some $W \in \cP(\cY|\cX)$ with $C(W) >0$, $\delta \in \bbR^+$ and $\epsilon \in (0,1)$. Consider a sequence of constant composition $(n,R_n)$ codes $\{ (f_n, \varphi_n)\}_{n \geq 1}$ with the common composition $Q_n \in \cS_3(\delta)$ and $R_n \eqdef C(W) + \sqrt{\frac{V_\epsilon(W)}{n}} \Phi^{-1}(\epsilon)$. For some $n_\mo(W, \epsilon, \delta) \in \bbZ^+$, we have 
\begin{equation}
\bar{\mP}_{\me}(f_n, \varphi_n) > \epsilon, \, \textnormal{ for all }  n \geq n_\mo(W, \epsilon, \delta). \label{eq:asym-lem1}
\end{equation}
\eot
 \label{lem:asym-lem1}
\end{lemma}

\begin{IEEEproof}
Define\footnote{Since $I(\cdot, W)$ is continuous over $\cP(\cX)$, $\gamma(\delta)$ is well-defined and positive.} $\gamma(\delta) \in \bbR^+$ as 
\begin{equation}
\gamma(\delta) \eqdef C(W) - \max_{Q \in \textnormal{cl}(\cS_3(\delta))}\mI(Q;W) .
\end{equation}
For any message $m$, let
\begin{equation}
G_n(m) \eqdef \left\{ \by^n : \frac{1}{n}\sum_{i=1}^n \ln \frac{W(Y_i|x_i(m))}{q_{Q_n}(Y_i)} > \mI(Q_n;W) + \frac{\gamma(\delta)}{2}\right\}.
\end{equation}
Define\footnote{Since $V(\cdot, W)$ is continuous over the compact set $\cP(\cX)$ (e.g., \cite[Lemma~62]{polyanskiy10}), $V_{\max}$ is well-defined and finite.} $V_{\max} \eqdef \max_{P \in \cP(\cX)}V(P,W) \in \bbR^+$.

The following arguments are essentially the ones used in \cite[Appendix~B]{altug12}, which we outline here for completeness. We have 
\begin{equation}
\bar{\mP}_{\me}(f_n, \varphi_n) = 1 - \frac{1}{|\cM_n|}\sum_{m \in \cM_n}\sum_{\by^n \in \cA_m \cap G_n(m)}W(\by^n|\bx^n(m)) - \frac{1}{|\cM_n|}\sum_{m \in \cM_n}\sum_{\by^n \in \cA_m \cap G_n^c(m)}W(\by^n|\bx^n(m)). \label{eq:asym-lem1-pf-0}
\end{equation}
Since $q_{Q_n}$ is a probability measure on $\cY^n$ and the decoding regions are disjoint, one can verify that 
\begin{equation}
\frac{1}{|\cM_n|}\sum_{m \in \cM_n}\sum_{\by^n \in \cA_m \cap G_n^c(m)}W(\by^n|\bx^n(m)) \leq e^{-n \left[ \frac{\gamma(\delta)}{2} + \sqrt{\frac{V_\epsilon(W)}{n}}\Phi^{-1}(\epsilon) \right]}. \label{eq:asym-lem1-pf-1}
\end{equation}
Moreover, via an application of Chebyshev's inequality, it is easy to verify that  
\begin{equation}
\frac{1}{|\cM_n|}\sum_{m \in \cM_n}\sum_{\by^n \in \cA_m \cap G_n(m)}W(\by^n|\bx^n(m)) \leq  \frac{nV(Q;W)}{\frac{(n\gamma(\delta))^2}{4}} \leq \frac{4 V_{\max}}{n \gamma(\delta)^2}. \label{eq:asym-lem1-pf-2}
\end{equation}
By plugging \eqref{eq:asym-lem1-pf-1} and \eqref{eq:asym-lem1-pf-2} into \eqref{eq:asym-lem1-pf-0} and choosing $n_\mo(W, \epsilon, \delta) \in \bbZ^+$ such that for all $n \geq n_\mo(W, \epsilon, \delta)$, we have  
\[
1 - e^{-n\left[ \frac{\gamma(\delta)}{2} + \sqrt{\frac{V_\epsilon(W)}{n}}\Phi^{-1}(\epsilon)\right]}- \frac{4 V_{\max}}{n \gamma(\delta)^2} > \epsilon,
\]
we conclude that \eqref{eq:asym-lem1} holds. 
\end{IEEEproof}

\begin{lemma}
Fix some $\epsilon \in (0.5,1)$, $W \in \cP(\cY|\cX)$ with $V_\epsilon(W) >0$, and $a \in \bbR^+$ with $a>\frac{2}{1-\epsilon}$. Consider an $(n,R_n)$ constant composition code $(f,\varphi)$ with $R_n = C(W) + \sqrt{\frac{V_\epsilon(W)}{n}}\Phi^{-1}(\epsilon) - \frac{\ln\left( 1 - \epsilon - \frac{2}{a}\right)}{n}$ and the common composition $Q$ satisfying $V(Q,W) < \frac{V_{\epsilon}(W)\left[\Phi^{-1}(\epsilon)\right]^2}{a}$. We have 
\begin{equation}
\bar{\mP}_{\me}(f, \varphi) > \epsilon.
\label{eq:asym-lem2}
\end{equation}
\eot
\label{lem:asym-lem2}
\end{lemma}

\begin{IEEEproof}
For any message $m$, define
\begin{equation}
G_n(m) \eqdef \left\{ \by^n : \frac{1}{n}\sum_{i=1}^n \ln \frac{W(Y_i|x_i(m))}{q_n(Y_i)} > C(W) + \sqrt{\frac{V_\epsilon(W)}{n}}\Phi^{-1}(\epsilon) \right\}. 
\end{equation}
Via arguments similar to the ones given in the proof of Lemma~\ref{lem:asym-lem1}, one can verify that 
\begin{align}
\bar{\mP}_{\me}(f, \varphi) & \geq 1 - \left( 1 - \epsilon - \frac{2}{a} \right) - \frac{nV(Q,W)}{\left[ n(C(W) - \mI(Q;W)) + \sqrt{n V_\epsilon(W)}\Phi^{-1}(\epsilon)\right]^2}  \\
& \geq \epsilon + \frac{1}{a} > \epsilon. 
\end{align}
\end{IEEEproof}

For any $Q \in \cP(\cX)$, define 
\begin{align}
U(Q,W) & \eqdef \sum_{x,y}Q(x)W(y|x)  \left[ \ln \frac{W(y|x)}{q_Q(y)} - \mI(Q;W) \right]^2. \\
m_3(Q,W) & \eqdef \sum_{x \in \cX}Q(x)\mE_{W(\cdot|x)}\left[ \left| \ln \frac{W(Y|x)}{q_Q(Y)} - \mE_{W(\cdot|x)}\left[ \ln \frac{W(Y|x)}{q_Q(Y)}\right] \right|^3\right].
\end{align}
Choose $\delta >0$ such that\footnote{Without loss of generality, we assume that $W$ has no all-zero column.} 
\begin{equation}
 \cS(q_Q) = \cY,  \textnormal{ for all } Q \in \cP(\cX) \backslash \cS_3(\delta). \label{eq:asym-delta} 
\end{equation}
Such a choice is possible due to the evident continuity of $\alpha_{y}(\cdot)$ for any $y \in \cY$ and the fact that the unique capacity achieving output distribution has full support, as noted before. The following has been shown by Polyanskiy \emph{et al.} \cite[Lemma~46]{polyanskiy10}
\begin{equation}
\tilde{m}_3(Q,W) \eqdef \sum_{(x,y)}Q(x)W(y|x)\left| \ln \frac{W(Y|X)}{q_Q(Y)} - \mI(Q;W) \right|^3 \leq \left(\frac{3\left( |\cX|^{1/3} + |\cY|^{1/3}\right)}{e} + \ln \min\{ |\cX|, |\cY| \}\right)^3 =: \kappa(W) \in \bbR^+. \label{eq:asym-alpha}
\end{equation}

Fix some $\nu \in \bbR^+$ and $\epsilon \in (0,1)$. Assume $\cS_1(\delta, \nu) \neq \emptyset$ and define\footnote{Since $m_3(\cdot,W)$ and $V(\cdot, W)$ are continuous over $\cP(\cX)$ (e.g., \cite[Lemma~62]{polyanskiy10}), $K(W, \epsilon, \delta, \nu)$ is well-defined, positive and finite.} 
\begin{equation}
K(W, \epsilon, \delta, \nu) \eqdef \frac{2}{\phi(\Phi^{-1}(\epsilon))}\left[ \max_{P \in \cS_1(\delta, \nu)}\frac{m_3(P,W)}{V(P,W)} + \left( \frac{1}{\sqrt{2\pi}} + \frac{\kappa(W)}{\nu}\right) \right]\in \bbR^+. \label{eq:asym-K}
\end{equation}

\begin{lemma}
Fix an asymmetric and singular $W \in \cP(\cY|\cX)$, $\epsilon \in (0, 1)$ and $\nu \in \bbR^+$. Choose $\delta \in \bbR^+$ such that \eqref{eq:asym-delta} holds. For some $\tilde{n}_\mo(W, \epsilon, \delta, \nu) \in \bbZ^+$ and any $n \geq \tilde{n}_\mo(W, \epsilon, \delta, \nu)$, consider an $(n,R_n)$ constant composition code $(f, \varphi)$ with common composition $Q \in \cS_1(\delta, \nu)$ and $R_n = \mI(Q;W) + \sqrt{\frac{V(Q,W)}{n}} \Phi^{-1}(\epsilon) + \frac{K(W, \epsilon, \delta, \nu)}{n}$. We have 
\begin{equation}
\bar{\mP}_{\me}(f, \varphi) >\epsilon.\label{eq:asym-lem3}
\end{equation}
\label{lem:asym-lem3}
\eot
\end{lemma}

\begin{IEEEproof}
Assume $\cS_1(\delta, \nu) \neq \emptyset$, because otherwise the claim is void. The proof is similar to the proof of Proposition~\ref{prop:sym}. Let $\tilde{n}_\mo(W, \epsilon, \delta, \nu) \in \bbZ^+$ be such that for all $n \geq \tilde{n}_\mo(W, \epsilon, \delta, \nu)$, 
\begin{equation}
\sqrt{n} > \frac{2 K(W, \epsilon, \delta, \nu)}{\phi(\Phi^{-1}(\epsilon))\sqrt{\nu}}.
\end{equation}
In light of \eqref{eq:asym-K}, the existence of such a choice is evident. 

Consider any $(n,R_n)$ constant composition code, say $(f, \varphi)$, with the common composition $Q$. Assume $Q$ and $R_n$ are as in the statement of the lemma. Consider any $\bx^n \in \cX^n$ and define 
\begin{equation}
M_{\bx^n}(\lambda) \eqdef \mE_{W(\cdot|\bx^n)}\left[ e^{\lambda \ln \frac{W(\bY^n|\bx^n)}{q_{P_{\bx^n}}(\bY^n)}}\right], \, \forall \, \lambda \in \bbR.
\end{equation}
We claim that for any $\bx^n, \bz^n \in \cX^n$ with $P_{\bx^n} = P_{\bz^n}$, we have 
\begin{equation}
M_{\bx^n}(\lambda) = M_{\bz^n}(\lambda), \, \forall \, \lambda \in \bbR. \label{eq:asym-lem3-pf0}
\end{equation}
To see this, simply note that 
\begin{align}
M_{\bx^n}(\lambda) & = \sum_{\by^n : W(\by^n | \bx^n) >0}e^{n \sum_{y}P_{\by^n}(y) \ln \delta_y} e^{-\lambda n \sum_{y}P_{\by^n}(y) \ln \alpha_y(P_{\bx^n})} \\
& = \sum_{P \in \cP_n(\cY)}e^{n \sum_{y}P(y) \ln \delta_y} e^{-\lambda n \sum_{y}P(y) \ln \alpha_y(P_{\bx^n})}|\{ \by^n : P_{\by^n} = P \textnormal{ and } W(\by^n |\bx^n) >0\}| \\
& = \sum_{P \in \cP_n(\cY)}e^{n \sum_{y}P(y) \ln \delta_y} e^{-\lambda n \sum_{y}P(y) \ln \alpha_y(P_{\bz^n})}|\{ \by^n : P_{\by^n} = P \textnormal{ and } W(\by^n |\bz^n) >0\}| \label{eq:asym-lem3-pf1}\\
& = M_{\bz^n}(\lambda), \label{eq:asym-lem3-pf1.1}
\end{align}
where \eqref{eq:asym-lem3-pf1} follows since $P_{\bx^n} = P_{\bz^n}$. Equation \eqref{eq:asym-lem3-pf0}, along with the uniqueness theorem for the moment generating functions (e.g., \cite[Ex.~26.7]{billingsley95}), and the fact that $q_Q$ is of full support, enables us to invoke Lemma~\ref{lem:multipurp} to deduce 
\begin{equation}
\bar{\mP}_{\me}(f, \varphi) \geq W(\cS_{R_n}(Q) | \bz^n) - \sum_{\by^n \in \cS_{R_n}(Q)}q_Q(\by^n) e^{-n\left[ R_n - \frac{1}{n}\sum_{i=1}^n \ln \frac{1}{\alpha_{y_i}(Q)}\right]}, \label{eq:asym-lem3-pf2}
\end{equation}
for a given $\bz^n \in \cX^n$ with $P_{\bz^n} = Q$. 

Due to the singularity of $W$, we have 
\begin{align}
W(\cS_{R_n}(Q) | \bz^n) & = \sum_{\by^n}W(\by^n | \bz^n)\b1\left\{ \frac{1}{n}\sum_{i=1}^n \ln \frac{W(y_i|z_i)}{q_Q(y_i)} \leq R_n \right\} \\
& \geq \epsilon - \frac{m_3(Q,W)}{\sqrt{n}V(Q,W)^{3/2}} + \frac{K(W, \epsilon, \delta, \nu) \phi(\Phi^{-1}(\epsilon))}{\sqrt{n V(Q,W)}}\left( 1 - \frac{K(W, \epsilon, \delta, \nu) }{2\sqrt{n V(Q,W)}\phi(\Phi^{-1}(\epsilon))}\right), \label{eq:asym-lem3-pf3}
\end{align}
where the proof of \eqref{eq:asym-lem3-pf3} is similar to that of \eqref{eq:sym-pf-final5} and omitted for brevity. 

Further, define $P_{X,Y}(x,y) \eqdef Q(x)W(y|x)$ and $P_{X,Y}(\bx^n, \by^n) \eqdef \prod_{i=1}^nP_{X,Y}(x_i, y_i)$. Evidently,  
    \begin{align}
     \sum_{\by^n \in \cS_{R_n}(Q)}q_Q(\by^n) e^{-n\left[ R_n - \frac{1}{n}\sum_{i=1}^n \ln \frac{1}{\alpha_{y_i}(Q)}\right]} & = \sum_{(\bx^n, \by^n)}P_{X,Y}(\bx^n, \by^n)\b1\left\{\frac{1}{n}\sum_{i=1}^n \ln \frac{W(y_i|x_i)}{q_Q(y_i)} \leq R_n \right\}e^{-n\left[R_n - \frac{1}{n}\sum_{i=1}^n \ln \frac{W(y_i|x_i)}{q_Q(y_i)}\right]} \\
     & \leq \frac{1}{\sqrt{2\pi n U(Q,W)}} + \frac{\tilde{m}_3(Q,W)}{\sqrt{n}U(Q,W)^{3/2}} \label{eq:asym-lem3-pf4}\\
     & \leq \frac{1}{\sqrt{n V(Q,W)}}\left( \frac{1}{\sqrt{2\pi}} + \frac{\kappa(W)}{V(Q,W)}\right), \label{eq:asym-lem3-pf5}
    \end{align}
where \eqref{eq:asym-lem3-pf4} follows from Lemma~\ref{lem:lem2}, whose application is ensured by the fact that $U(Q,W) \geq V(Q,W)$ (e.g., \cite[Lemma~62]{polyanskiy10}), which, along with \eqref{eq:asym-alpha}, also implies \eqref{eq:asym-lem3-pf5}.  

By plugging \eqref{eq:asym-lem3-pf3} and \eqref{eq:asym-lem3-pf5} into \eqref{eq:asym-lem3-pf2}, along with the definitions of $K(W,\epsilon, \delta, \nu)$ and $n_\mo(W, \epsilon, \delta, \nu)$, one can verify that 
\begin{equation}
\bar{\mP}_{\me}(f,\varphi) > \epsilon + \frac{1}{\sqrt{n V(Q,W)}}\left( \max_{P \in \cS_1(\delta,\nu)} \frac{m_3(P,W)}{V(P,W)} - \frac{m_3(Q,W)}{V(Q,W)}\right) \geq \epsilon,
\end{equation}
which, in turn, implies \eqref{eq:asym-lem3}.
\end{IEEEproof}

To prove item (i) of Proposition~\ref{prop:asym}, fix some $\epsilon \in (0,0.5)$ and assume $V_\epsilon(W) >0$, because otherwise \cite[Proposition~9]{tomamichel-tan13} implies \eqref{eq:prop-asym-1}. Fix some $\delta >0$ such that \eqref{eq:asym-delta} holds and $\cS_2\left(\delta, \frac{V_\epsilon(W)}{2}\right) = \emptyset$. Such a choice is possible since $V(\cdot,W)$ is continuous over $\cP(\cX)$, as noted before. For any $P \in \cP(\cX)$, let $P^{\ast}(P) \eqdef \arg \min_{Q \in \cP^\ast_W}|| Q - P||_2$. Fix some $\beta_1, \beta_2 \in \bbR^+$ such that 
\begin{equation}
\mI(P;W) \leq C(W) - \beta_1||P - P^\ast(P) ||_2^2, \quad |\sqrt{V(P,W)} - \sqrt{V(P^\ast(P),W)}| \leq \beta_2 ||P - P^\ast(P) ||_2,
\label{eq:asym-beta}
\end{equation}
for any $P \in \cS_1\left(\delta, \frac{V_\epsilon(W)}{2}\right)$, whose existence is ensured by \cite[Lemma~7]{tomamichel-tan13}. In light of \eqref{eq:asym-beta}, for all $P \in \cS_1\left(\delta, \frac{V_\epsilon(W)}{2}\right)$ and for any $n \in \bbZ^+$, 
\begin{align}
n\mI(P;W) + \sqrt{n V(P,W)}\Phi^{-1}(\epsilon) & \leq nC(W) + \sqrt{n V_\epsilon(W)}\Phi^{-1}(\epsilon) \nonumber \\
& \quad  -\beta_1  n ||P - P^\ast(P) ||_2^2 + \beta_2 |\Phi^{-1}(\epsilon)|\sqrt{n}||P - P^\ast(P) ||_2  \\
& \leq nC(W) + \sqrt{n V_\epsilon(W)}\Phi^{-1}(\epsilon) + \frac{\left(\beta_2 |\Phi^{-1}(\epsilon)|\right)^2}{4 \beta_1}, \label{eq:asym-final1}
\end{align}
where \eqref{eq:asym-final1} follows from elementary calculus. Consider any $n \in \bbZ^+$ such that 
\begin{equation}
n \geq \max\{n_\mo(W, \epsilon, \delta), \tilde{n}_\mo(W, \epsilon, \delta, V_\epsilon(W)/2)\},
\end{equation}
where $n_\mo$ and $\tilde{n}_\mo$ are as given in Lemmas~\ref{lem:asym-lem1} and \ref{lem:asym-lem3}, respectively. Define 
\begin{equation}
R_n \eqdef C(W) + \sqrt{\frac{V_\epsilon(W)}{n}}\Phi^{-1}(\epsilon) + \frac{\frac{\left(\beta_2 |\Phi^{-1}(\epsilon)|\right)^2}{4 \beta_1} + K(W, \epsilon, \delta, V_\epsilon(W)/2)}{n},
\end{equation}
and consider any $(n,R_n)$ constant composition code $(f, \varphi)$ with the common composition $Q$. Now, if $Q \in \cS_3(\delta)$, then Lemma~\ref{lem:asym-lem1} implies that $\bar{\mP}_{\me}(f, \varphi) >\epsilon$. Similarly, if $Q \in \cS_1\left(\delta, \frac{V_\epsilon(W)}{2}\right) $, then Lemma~\ref{lem:asym-lem3} and \eqref{eq:asym-final1} implies that $\bar{\mP}_{\me}(f, \varphi) >\epsilon$. Since the code is arbitrary, we conclude the proof of item (i) of the proposition. 

To prove item (ii) of Proposition~\ref{prop:asym}, fix some $\epsilon \in (0.5, 1)$ and $\delta >0$ such that \eqref{eq:asym-delta} holds. Choose some $a \in \bbR^+$ that satisfies $a > \frac{2}{1-\epsilon}$ and $\nu \in \bbR^+$ such that $\nu \leq \frac{V_\epsilon(W)\left[\Phi^{-1}(\epsilon)\right]^2}{a}$. Similar to \eqref{eq:asym-beta}, choose $\beta_1, \beta_2 \in \bbR^+$ such that 
\begin{equation}
\mI(P;W) \leq C(W) - \beta_1||P - P^\ast(P) ||_2^2, \quad |\sqrt{V(P,W)} - \sqrt{V(P^\ast(P),W)}| \leq \beta_2 ||P - P^\ast(P) ||_2,
\label{eq:asym-beta-1}
\end{equation}
for any $P \in \cS_1\left(\delta, \nu \right)$. From \eqref{eq:asym-beta-1}, similar to \eqref{eq:asym-final1}, we deduce that for all $P \in \cS_1(\delta, \nu)$ and $n \in \bbZ^+$, 
\begin{equation}
n\mI(P;W) + \sqrt{n V(P,W)}\Phi^{-1}(\epsilon) \leq  nC(W) + \sqrt{n V_\epsilon(W)}\Phi^{-1}(\epsilon) + \frac{\left(\beta_2 \Phi^{-1}(\epsilon)\right)^2}{4 \beta_1}. \label{eq:asym-final2}
\end{equation}
Consider any $n \in \bbZ^+$ such that $n \geq \max\{  n_\mo(W, \epsilon, \delta),   \tilde{n}_\mo(W, \epsilon, \delta, \nu) \}$, where $n_\mo$ and $\tilde{n}_\mo$ are as given in Lemmas~\ref{lem:asym-lem1} and \ref{lem:asym-lem3}, respectively. Consider any $(n,R_n)$ constant composition code $(f, \varphi)$ with the common composition $Q$ and define 
\begin{equation}
R_n \eqdef C(W) + \sqrt{\frac{V_\epsilon(W)}{n}}\Phi^{-1}(\epsilon) + \frac{\frac{\left(\beta_2 \Phi^{-1}(\epsilon)\right)^2}{4 \beta_1} + K(W, \epsilon, \delta, \nu) - \ln \left(1-\epsilon- \frac{2}{a}\right) }{n}.
\end{equation}
If $Q \in \cS_3(\delta)$, then $\bar{\mP}_\me(f, \varphi) > \epsilon$ due to Lemma~\ref{lem:asym-lem1}. If $Q \in \cS_2(\delta, \nu)$, then $\bar{\mP}_\me(f, \varphi) > \epsilon$ because of Lemma~\ref{lem:asym-lem2}. Finally, if $Q \in \cS_1(\delta, \nu)$, then Lemma~\ref{lem:asym-lem3}, along with \eqref{eq:asym-final2}, implies that $\bar{\mP}_\me(f, \varphi) > \epsilon$. Since the code is arbitrary, we conclude that \eqref{eq:prop-asym-2} holds. \qed

\section{Discussion}
\label{sec:discussion}
\subsection{Relation to the minimax converse}
\label{ssec:minmax}
One can interpret the arguments leading to the proof of Proposition~\ref{prop:sym} in terms of the minimax converse (e.g., \cite[Theorem~1]{polyanskiy13}), which we illustrate next. To this end, we fix a symmetric and singular $W \in \cP(\cY|\cX)$ and note that \cite[Eq.~(9) and (11)]{polyanskiy13} imply that for any $n \in \bbZ^+$ and $\epsilon \in (0,1)$,
\begin{equation}
M^\ast(n, \epsilon) \leq \frac{1}{\min_{P_{\bX^n}} \max_{Q_{\bY^n}} \beta_{1-\epsilon}(P_{\bX^n, \bY^n}, P_{\bX^n} \times Q_{\bY^n})}, \label{eq:conc-minmax-1}
\end{equation}
where $P_{\bX^n, \bY^n}(\bx^n, \by^n) \eqdef P_{\bX^n}(\bx^n)W(\by^n|\bx^n)$, $(P_{\bX^n} \times Q_{\bY^n})(\bx^n, \by^n) \eqdef P_{\bX^n}(\bx^n)Q_{\bY^n}(\by^n)$ and $\beta_{1-\epsilon}(P_{\bX^n, \bY^n}, P_{\bX^n} \times Q_{\bY^n})$ denotes the minimum probability of error under $P_{\bX^n} \times Q_{\bY^n}$, subject to the constraint that the error probability under hypothesis $P_{\bX^n, \bY^n}$ does not exceed $\epsilon$. Due to \cite[Theorem~21]{polyanskiy13}, the minimum on the right side of \eqref{eq:conc-minmax-1} is attained by $U_{\cX^n}$. Consider some $n \in \bbZ^+$ such that \eqref{eq:sym-pf-final1} holds and let $R$ be as in \eqref{eq:sym-pf-final2}. With these choices, we define\footnote{The following non-product distribution is inspired by \cite[Eq.~(168)]{polyanskiy13}. In particular, if $W$ is BEC then \eqref{eq:conc-minimax-2} reduces to \cite[Eq.~(168)]{polyanskiy13}.}
\begin{equation}
Q^\ast_{\bY^n}(\by^n) \eqdef \frac{e^{n \sum_{y}P_{\by^n}(y) \ln \delta_y } \b1\left\{ \by^n \in \cS_R \right\}}{\sum_{\bb^n}e^{n \sum_{b}P_{\bb^n}(b) \ln \delta_b } \b1\left\{ \bb^n \in \cS_R \right\}},
\label{eq:conc-minimax-2}
\end{equation}
where $\delta_y$ and $\cS_R$ are as defined before. Evidently, $Q^\ast_{\bY^n} \in \cP(\cY^n)$. With a slight abuse of notation, let $\beta_{1 - \epsilon}(U_{\cX^n}, Q^\ast_{\bY^n})$ denote the cost function of the optimization problem in the denominator of \eqref{eq:conc-minmax-1} when $P_{\bX^n} = U_{\cX^n}$ and $Q_{\bY^n} = Q^\ast_{\bY^n}$. Evidently, 
\begin{equation}
M^\ast(n, \epsilon) \leq \frac{1}{\beta_{1 - \epsilon}(U_{\cX^n}, Q^\ast_{\bY^n})}. 
\label{eq:conc-minimax-3}
\end{equation}
From the Neyman-Pearson Lemma (e.g., \cite{neyman-pearson33}), the right side of \eqref{eq:conc-minimax-3} is attained by a randomized threshold test with the randomization parameter $\tau \in (0,1)$ that satisfies
\begin{equation}
\tau W(\cS_R | \bx^n_\mo) = \epsilon \quad \textnormal{ and } \quad \beta_{1 - \epsilon}(U_{\cX^n}, Q^\ast_{\bY^n}) = \frac{(1-\tau)W(\cS_R | \bx^n_\mo)}{e^{nR} \sum_{\by^n \in \cS_R}q(\by^n)e^{-n\left[R - \frac{1}{n}\sum_{i=1}^n \ln \frac{1}{\alpha_{y_i}}\right]}} \label{eq:conc-minimax-4}. 
\end{equation}
Equation \eqref{eq:conc-minimax-4} can be verified via elementary algebra by noticing the fact that $W$ is singular and symmetric, and we omit the details for brevity. Finally,  \eqref{eq:sym-pf-final4} and \eqref{eq:sym-pf-final5}, along with \eqref{eq:sym-pf-final1} and \eqref{eq:sym-pf-final2}, imply that 
\begin{equation}
W(\cS_R|\bx^n_\mo) - \sum_{\by^n \in \cS_R}q(\by^n)e^{-n\left[R - \frac{1}{n}\sum_{i=1}^n \ln \frac{1}{\alpha_{y_i}}\right]} > \epsilon. \label{eq:conc-minimax-5} 
\end{equation}
Equations \eqref{eq:conc-minimax-3}, \eqref{eq:conc-minimax-4} and \eqref{eq:conc-minimax-5} imply that $M^\ast(n, \epsilon) < e^{nR}$, which, in turn, implies Proposition~\ref{prop:sym}.

In light of the above discussion, the proof of Proposition~\ref{prop:sym} would be shorter had we used the minimax converse with the output distribution given in \eqref{eq:conc-minimax-2}. However, we opt to use Lemma~\ref{lem:multipurp} because it makes the role of $\cS_R$ more transparent, as explained in Remark~\ref{rem:lemma1}. 

\subsection{On dropping the constant composition assumption}
\label{ssec:CCC}
As noted before, Proposition~\ref{prop:asym} gives an $O(1)$ upper bound on the third-order term of the normal approximation for asymmetric and singular DMCs only if we consider constant composition codes. Although this restriction is undesirable, it is quite common in converse results. Indeed, the usual proof of the converse statement of \eqref{eq:intro-1} involves first showing it for constant composition codes, and then arguing that this restriction at most results in an extra $O(\ln n)$ term. 

It should be noted that if the channel has sufficient symmetry, then the constant composition step is not necessary and one can derive an $\ln \sqrt{n}$ upper bound on the third-order term \cite[Sec.~3.4.5]{polyanskiy10-phd}. Recently, Tomamichel-Tan~\cite{tomamichel-tan13} have showed an $\ln \sqrt{n} $ upper bound on the third-order term in general by dispensing with the constant composition code restriction in the first step. This result, coupled with the existing results in the literature, gives the third-order term for a broad class of channels, which includes positive channels with positive capacity \cite{tomamichel-tan13}, but does not include asymmetric and singular channels. The method of \cite{tomamichel-tan13} is essentially based on relating the channel coding problem to a binary hypothesis test by using an auxiliary output distribution, which is in the same vein as the so-called meta-converse of Polyanskiy \emph{et al.} (e.g., \cite[Section~III.E and III.F]{polyanskiy10}). As opposed to the classical applications of this idea, which use a product auxiliary output distribution and result in the aforementioned two-step procedure, the authors of \cite{tomamichel-tan13} uses an appropriately chosen non-product output distribution to dispense with the constant composition step. However, their non-product distribution is different from the one used in the previous subsection and it is an interesting future research topic to investigate how to combine the analysis of \cite{tomamichel-tan13} and the viewpoint in Section~\ref{ssec:minmax} to drop the constant composition assumption in Proposition~\ref{prop:asym}. 

\subsection{Limitation in the error exponents regime}
\label{ssec:EA}
 One might conjecture that by following the same program used to prove Proposition~\ref{prop:asym}, one could prove the following lower bound for asymmetric and singular channels 
\begin{equation}
\liminf_{n \rightarrow \infty}\frac{\bar{\mP}_{\me, c}(n,R)}{\frac{e^{-n \mE_{\mSP}(R,W)}}{\sqrt{n}}} 
 \geq K(R,W), \label{eq:conc-EA}
\end{equation}
where $K(R,W)$ is a positive constant that depends on $R$ and $W$, and $\mE_{\mSP}(R,W)$ is the sphere-packing exponent (e.g., \cite[Eq.~(5.8.2)]{gallager68}) 
\begin{equation}
\mE_{\mSP}(R,W) \eqdef \max_{Q \in \cP(\cX)} \mE_{\mSP}(R,Q,W), \, \mE_{\mSP}(R,Q,W) \eqdef \sup_{\rho \geq 0} \left\{ -\rho R - \ln \sum_{y \in \cY}\left( \sum_{x \in \cX} Q(x)W(y|x)^{1/(1+\rho)}\right)^{(1+\rho)}\right\}.
\end{equation}
However, a proof of \eqref{eq:conc-EA} seems to be more involved than its counterpart in the normal approximation regime, i.e., Proposition~\ref{prop:asym}. The main technical difficulty is proving the continuity properties of $\mE_{\mSP}(R,\cdot,W)$ that are required to distinguish between the ``good types'', for which $\mE_{\mSP}(R,Q,W) \approx \mE_{\mSP}(R,W)$ and hence one can use a result like Lemma~\ref{lem:asym-lem3} to deduce an $\Omega(1/\sqrt{n})$ sub-exponential term directly, and the ``bad types'', for which $\mE_{\mSP}(R,Q,W)$ is bounded away from $\mE_{\mSP}(R,W)$ and hence one can utilize this inferiority of the exponent to deduce an $\Omega(1/\sqrt{n})$ sub-exponential term. Indeed, the proof of these continuity properties appears to be quite intricate.

\section*{acknowledgment}
The authors thank Sergio Verd\'u for raising the question of whether the
technique used in~\cite{altug13} could be applied to the normal approximation
regime.
The first author thanks Paul Cuff for his hospitality while portions of this work were being completed during his visit to Princeton University. This research is supported by the National Science Foundation under grant CCF-1218578.



\begin{thebibliography}{99}

\bibitem{juschkewtisch53}
A.~A.~Yushkevich, ``On limit theorems connected with the concept of entropy of Markov chains,'' (in Russian) \emph{Uspekhi Mat. Nauk}, vol.~8, no.~5(57), pp.~177--180, 1953. 

\bibitem{strassen62}
V.~Strassen, ``Asymptotische absch\"{a}tzungen in Shannon's informationstheorie,'' \emph{Trans. Third Prague Conf. Information Theory}, 1962,
Czechoslovak Academy of Sciences, Prague, pp. 689-723.

\bibitem{hayashi09} 
M.~Hayashi, ``Information spectrum approach to second-order coding rate in channel coding,'' \emph{IEEE Trans. on Information Theory}, vol.~IT~55, no.~11, pp.~4947--4966, Nov.~2009.

\bibitem{Wang09}
L.~Wang, R.~Renner, and R.~Colbeck, ``Simple channel coding bounds,'' in \emph{Proc. 2009 IEEE Int. Symp. Inf. Theory}, Seoul, S.~Korea, Jul.~2009, pp.~1804--1808. 

\bibitem{polyanskiy09-1}
Y.~Polyanskiy, H.~V.~Poor and S.~Verd\'u, ``Dispersion of Gaussian channels,'' in \emph{Proc. 2009 IEEE Int. Symp. Inf. Theory}, Seoul, S.~Korea, Jul.~2009, pp.~2204--2208. 

\bibitem{polyanskiy10}
Y.~Polyanskiy, H.~V.~Poor and S.~Verd\'u, ``Channel coding rate in the finite blocklength regime,'' \emph{IEEE Trans. on Information Theory}, vol.~IT--56, no.~5, pp.~2307--2359, May~2010.

\bibitem{polyanskiy10-phd}
Y.~Polyanskiy, ``Channel coding: non-asymptotic fundamental limits,'' Ph.D. dissertation, Princeton Univ., Princeton, NJ, Nov.~2010. 

\bibitem{polyanskiy10-1}
Y.~Polyanskiy and S.~Verd\'u, ``Arimoto channel coding converse and R\'enyi divergence,'' in \emph{Proc. 48th Annual Allerton Conf. Communications, Control, and Computing}, Monticello, IL, Oct.~2010, pp.~1327--1333.

\bibitem{polyanskiy11}
Y.~Polyanskiy, H.~V.~Poor and S.~Verd\'u, ``Dispersion of the Gilbert-Elliott channel,'' \emph{IEEE Trans. on Information Theory}, vol.~IT--57, no.~4, pp.~1829--1848, Apr.~2011.

\bibitem{martinez11}
A.~Martinez and A.~Guill\'en i F\`abregas, ``Random-coding bounds for threshold decoders: error exponent and saddlepoint approximation,'' in \emph{Proc. 2011 IEEE Int. Symp. Inf. Theory}, St.~Petersburg, Russia, Aug.~2011, pp.~2899--2903.   

\bibitem{polyanskiy11-1}
Y.~Polyanskiy and S.~Verd\'u, ``Scalar coherent fading channel: dispersion analysis,'' in \emph{Proc. 2011 IEEE Int. Symp. Inf. Theory}, St.~Petersburg, Russia, Aug.~2011, pp.~2959--2963.

\bibitem{chen11}
P.~N.~Chen, H.~Y.~Lin and S.~M.~Moser, ``Ultra-small block-codes for binary discrete memoryless channels,'' in \emph{Proc. IEEE Inf. Theory Workshop}, Paraty, Brazil, Oct.~2011, pp.~175--179. 

\bibitem{riedl11}
T.~J.~Riedl, T.~J.~Coleman and A.~C.~Singer ``'Finite block-length achievable rates for queuing timing channels,'' in \emph{Proc. IEEE Inf. Theory Workshop}, Paraty, Brazil, Oct.~2011, pp.~200--204. 

\bibitem{hoydis12}
J.~Hoydis, J.~Couillet, R.~Piantanida and M.~Debbah, ``A random matrix approach to the finite blocklength regime of MIMO fading channels,'' in \emph{Proc. 2012 IEEE Int. Symp. Inf. Theory}, Boston, MA, Jul.~2012, pp.~2181--2185. 

\bibitem{moulin12}
P.~Moulin, ``The log-volume of optimal constant-composition codes for memoryless channels, within O(1) bits,'' in \emph{Proc. 2012 IEEE Int. Symp. Inf. Theory}, Boston, MA, Jul.~2012, pp.~826--830.

\bibitem{varshney12}
L.~R.~Varshney, S.~J.~Mitter and V.~K.~Goyal, ``An information-theoretic characterization of channels that die,'' \emph{IEEE Trans. on Information Theory}, vol.~IT--58, no.~9, pp.~5711--5724, Sep.~2012. 

\bibitem{yang12}
W.~Yang, W.~Durisi, G.~Koch and Y.~Polyanskiy, ``Diversity versus channel knowledge at finite block-length,'' in \emph{Proc. IEEE Inf. Theory Workshop}, Lausanne, Switzerland, Sep.~2012, pp.~572--576. 

\bibitem{polyanskiy12}
Y.~Polyanskiy and S.~Verd\'u, ``Empirical distribution of good channel codes with non-vanishing error probability,'' submitted to \emph{IEEE Trans. on Information Theory}. Available from: \url{http://people.lids.mit.edu/yp/homepage/data/optcodes_journal.pdf}

\bibitem{polyanskiy13-1}
Y.~Polyanskiy and Y.~Wu, ``Peak-to-average power ratio of good codes for Gaussian channel,'' submitted to \emph{IEEE Trans. on Information Theory}. Available from: \url{http://arxiv.org/pdf/1302.0084v1.pdf}

\bibitem{ingber13}
A.~Ingber, R.~Zamir and M.~Feder, ``Finite-dimensional infinite constellations,'' \emph{IEEE Trans. on Information Theory}, vol.~IT--59, no.~3, pp.~1630--1656, Mar.~2013.  

\bibitem{Scarlett13}
J.~Scarlett, A.~Martinez, A.~Guill\`en i F\'abregas, ``Mismatched decoding: finite-length bounds, error exponents and approximations,'' submitted to \emph{IEEE Trans. on Information Theory}. Available from: \url{http://arxiv.org/pdf/1303.6166v1.pdf}

\bibitem{polyanskiy13}
Y.~Polyanskiy, ``Saddle point in the minimax converse for channel coding,'' \emph{IEEE Trans. on Information Theory}, vol.~IT--59, no.~5, pp.~2576--2595, May~2013.

\bibitem{haim13}
E.~Haim, Y.~Kochman and U.~Erez, ``The importance of tie-breaking in finite-blocklength bounds,'' in \emph{Proc. 2013 IEEE Int. Symp. Inf. Theory}, \.Istanbul, Turkey, Jul.~2013.

\bibitem{hoydis13}
J.~Hoydis, R.~Couillet and P.~Piantanida, ``Bounds on the second-order coding rate of the MIMO Rayleigh block-fading channel,'' in \emph{Proc. 2013 IEEE Int. Symp. Inf. Theory}, \.Istanbul, Turkey, Jul.~2013. 

\bibitem{kostina13}
V.~Kostina and S.~Verd\'u, ``Channels with cost constraints: strong converse and dispersion,'' in \emph{Proc. 2013 IEEE Int. Symp. Inf. Theory}, \.Istanbul, Turkey, Jul.~2013. 

\bibitem{moulin13}
P.~Moulin, ``Asymptotic Neyman-Pearson games for converse to the channel coding theorem,'' in \emph{Proc. 2013 IEEE Int. Symp. Inf. Theory}, \.Istanbul, Turkey, Jul.~2013.

\bibitem{raginsky13}
M.~Raginsky and I.~Sason, ``Refined bounds on the empirical distribution of
good channel codes via concentration inequalities,'' in \emph{Proc. 2013 IEEE Int. Symp. Inf. Theory}, \.Istanbul, Turkey, Jul.~2013. 

\bibitem{shkel13}
Y.~Y.~Shkel, V.~Y.~F.~Tan and S.~C.~Draper, ``Converse bounds for assorted codes in the finite blocklength regime,'' in \emph{Proc. 2013 IEEE Int. Symp. Inf. Theory}, \.Istanbul, Turkey, Jul.~2013. 

\bibitem{vazquez-vilar13}
G.~Vazquez-Vilar, A.~Tauste Campo, A.~Guill\`en i F\'abregas and A.~Martinez, ``The meta-converse bound is tight,''  in \emph{Proc. 2013 IEEE Int. Symp. Inf. Theory}, \.Istanbul, Turkey, Jul.~2013. 

\bibitem{tomamichel13-1}
M.~Tomamichel and V.~Y.~F.~Tan, ``$\epsilon$-capacities and second-order coding rates for channels with general state,'' submitted to \emph{IEEE Trans. on Information Theory}. Available from: \url{http://arxiv.org/pdf/1305.6789v1.pdf}

\bibitem{yang13}
W.~Yang, G.~Durisi, T.~Koch and Y.~Polyanskiy, ``Quasi-static SIMA fading channels at finite blocklength,'' in \emph{Proc. 2013 IEEE Int. Symp. Inf. Theory}, \.Istanbul, Turkey, Jul.~2013. 

\bibitem{tomamichel-tan13}
M.~Tomamichel and V.~Y.~F.~Tan, ``A tight upper bound for the third-order asymptotics for most discrete memoryless channels,'' to appear \emph{IEEE Trans. on Information Theory}.

\bibitem{altug13}
Y.~Altu\u{g} and A.~B.~Wagner, ``Exact asymptotics of the error probability in channel coding: symmetric channels,'' to be submitted to \emph{IEEE Trans. on Information Theory}. 

\bibitem{csiszar-korner81}
I.~Csisz\'ar and J.~K\"{o}rner, \emph{Information Theory: Coding Theorems for Discrete Memoryless Systems}. New York: Academic Press, 1981. 

\bibitem{gallager68}
R.~G.~Gallager, \emph{Information Theory and Reliable Communication}. New York: Wiley, 1968.

\bibitem{altug13-1}
Y.~Altu\u{g} and A.~B.~Wagner, ``Refinement of the random coding bound,'' to be submitted to \emph{IEEE Trans. on Information Theory}. 

\bibitem{korolev2010}
V.~Yu.~Korolev and I.~G.~Shevtsova, ``A new moment-type estimate of convergence rate in the Lyapunov theorem,''
\emph{Theory Probab. Appl.} vol.~55, no.~3, pp.~505--509, 2011. 

\bibitem{billingsley95}
P.~Billingsley, \emph{Probability and measure, 3rd edition}. Hoboken, NJ: Wiley, 1995. 

\bibitem{altug12}
Y.~Altu\u{g} and A.~B.~Wagner, ``Moderate deviation analysis of channel coding,'' submitted to \emph{IEEE Trans. on Information Theory}. Available from: \url{http://arxiv.org/pdf/1208.1924v1.pdf}

\bibitem{neyman-pearson33}
J.~Neyman and E.~S.~Pearson, ``On the problem of the most efficient tests of statistical hypothesis,'' \emph{Philos. Trans. Roy. Soc. London. Ser.~A}, vol.~231, pp.~289--337, 1933. 

\bibitem{SGB67}
C.~E.~Shannon, R.~G.~Gallager and E.~R.~Berlekamp, ``Lower bounds to error probability for coding on discrete memoryless channels,'' \emph{Inform. Contr.}, vol.~10, pp.~65--103, Jan. 1967

\bibitem{altug12-SP}
Y.~Altu\u{g} and A.~B.~Wagner, ``Refinement of the sphere-packing bound: asymmetric channels,'' submitted to \emph{IEEE Trans. on Information Theory}. Available from: \url{http://arxiv.org/pdf/1211.6697v1.pdf}

\end{thebibliography}
\end{document}